\documentclass[a4paper,11pt]{article}
\usepackage{amsmath, amssymb, graphics, graphicx}
\usepackage{color}
\usepackage{physics}
\usepackage{longtable}
\usepackage{subfigure}
\usepackage{multirow}
\usepackage{slashed}
\usepackage{mathtools}
\usepackage{hyperref}
\usepackage{cite}

\usepackage{stackengine}
\stackMath

\newcommand{\mathsym}[1]{{}}

\newcommand{\rmd}{\mathrm{d}}
\newcommand{\rmi}{\mathrm{i}}
\newcommand{\rme}{\mathrm{e}}

\allowdisplaybreaks

\usepackage[shortlabels]{enumitem}

\begin{document}


\begin{titlepage}
\begin{center}
\vspace*{-1.0cm}

\vspace{2.0cm}

{\LARGE  {\fontfamily{lmodern}\selectfont \bf The General Supersymmetric Solution \\[1ex] of Minimal Massive Supergravity}} \\[.2cm]

\vskip 1.5cm
{\large {\bf Nihat Sadik Deger$^{a,b,f}$, Jan Rosseel$^{c,f}$ and Henning Samtleben$^{d,e}$}}\\
\vskip 1.2cm

\begin{small}
{}$^a$ \textit{Department of Mathematics, Bogazici University, Bebek, 34342, Istanbul, T\"urkiye}

\vspace{5mm}

{}$^b$ \textit{Feza Gursey Center for Physics and Mathematics, Bogazici University, Kandilli, 34684,\\
Istanbul, Türkiye}

\vspace{5mm}

{}$^c$ \textit{Division of Theoretical Physics, Rudjer Bo\v{s}kovi\'c Institute, \\
Bijeni\v{c}ka 54, 10000 Zagreb, Croatia} \\

\vspace{5mm}

{}$^d$ \textit{ENSL, CNRS, Laboratoire de physique, F-69342 Lyon, France}

\vspace{5mm}

{}$^e$ \textit{Institut Universitaire de France (IUF)}

\vspace{5mm}

{}$^f$ \textit{Erwin Schr\"odinger International Institute for Mathematics and Physics, \\ University of Vienna, Boltzmanngasse 9, 1090, Vienna, Austria}

\vspace{4mm}

{\texttt{sadik.deger@bogazici.edu.tr, Jan.Rosseel@irb.hr, henning.samtleben@ens-lyon.fr}}

\end{small}

\end{center}

\vskip 1.4 cm
\begin{abstract}
\vskip.2cm\noindent
Minimal massive supergravity is the supersymmetric extension of
minimal massive gravity in three dimensions. 
The theory admits a supersymmetric 
anti-de Sitter vacuum around which the propagating modes of spin 2 and 3/2 combine into a supermultiplet.
In this letter, we determine the most general supersymmetric solution of the theory by analyzing its Killing spinor equations. Just as for topologically massive supergravity, the general supersymmetric solution is a plane wave with a null Killing vector field. 
As a particular subclass we find null-warped AdS$_3$ spaces
and, with proper periodic identifications, null-warped AdS$_3$ black holes.

\end{abstract}

\end{titlepage}


%

\section{Introduction}

Minimal massive gravity (MMG) in three dimensions is an intriguing example of massive gravity~\cite{Bergshoeff:2014pca}.
It is a deformation of topologically massive gravity (TMG) \cite{Deser:1982vy,Deser:1981wh} by a term quadratic in the Schouten tensor. The model still admits an Anti-de Sitter (AdS) vacuum, but as a consequence of the deformation it overcomes the unitarity problems from which most other 3D massive gravity models suffer: there is a region in parameter space of MMG in which the massive spin 2 mode around the AdS vacuum is neither a ghost nor tachyonic, while both Brown-Henneaux central charges remain positive. That is, both bulk and boundary theory are unitary.
The supersymmetric extension to minimal massive supergravity (MMSG) has been constructed recently \cite{Deger:2022gim,Deger:2023eah}. In particular, it was found that every MMG model carrying unitary AdS vacua admits a supersymmetric extension in which these AdS vacua are supersymmetric. Accordingly, the propagating modes of spin 2 and 3/2 around the vacuum combine into an AdS supermultiplet of the dual superconformal field theory.

In this letter, we give a systematic study of the Killing spinor equations of MMSG in order to identify its most general supersymmetric solution. The analysis closely follows the lines of \cite{Gibbons:2008vi} for topologically massive supergravity (TMSG) \cite{Deser:1982sw,Deser:1982sv} and interestingly reduces to the same type of differential equation. Accordingly, the supersymmetric solutions of MMSG are in correspondence with the supersymmetric solutions of TMSG and given by plane waves with a null Killing vector field. A particular subclass thereof corresponds to null warped AdS$_3$ spaces. Such solutions have appeared before in the analysis of supersymmetric solutions of ${\cal N}=(1,1)$ and ${\cal N}=(2,0)$ TMSG \cite{Deger:2013yla,Deger:2016vrn}, respectively, as well as among the supersymmetric solutions of new massive gravity \cite{Alkac:2015lma}, general massive gravity \cite{Deger:2018kur}, and ${\cal N}=(4,0)$, ${\rm SO}(4)$ gauged supergravity \cite{Deger:2024xnd}.

The rest of this letter is organized as follows. In section~2 we review the structure of MMG and its supersymmetric extensions. In section~3, we analyze its Killing spinor equations and derive the general supersymmetric solution which is a plane wave with a null Killing vector field. In turn, we find that any such solution is supersymmetric in one of the supersymmetric extensions of MMG. A particular subclass yields null-warped AdS$_3$ spaces. In section 4, we discuss the resulting null-warped AdS$_3$ black holes upon proper periodic identifications.

\section{Minimal Massive Supergravity}

A convenient manner to formulate minimal massive supergravity (MMSG) is in terms of a dreibein $e_{\mu}{}^a$, a one-form $\varpi_{\mu}{}^{a}$ and two gravitino fields $\psi_\mu$ and $\Psi_\mu$. Its Lagrangian is up to an overall factor explicitly given by \cite{Deger:2022gim}
\begin{align}
{\cal L}[e,\varpi,\psi,\Psi] =\;&
\varepsilon^{\mu\nu\rho}\Big(e_\mu{}^{a}R[\omega]_{\nu\rho, a}
+\lambda\, \varepsilon_{a b c} e_\mu{}^{a} e_\nu{}^{b} e_\rho{}^{c}
+ \tfrac{\tau}{2}\,e_\mu{}^{a}{} T[\varpi]_{\nu\rho a} \Big)
+\kappa\,{\cal L}_{\rm CS}[\varpi] 
\nonumber\\
&{}\,
-\varepsilon^{\mu\nu\rho}
\bar\psi{}_\mu D[\omega]_\nu \psi{}_\rho
  +\frac{1}4\left(\eta\,\tau+\tfrac1{\eta\,\kappa}\right)
  \varepsilon^{\mu\nu\rho}
 \,\bar\psi{}_\mu \gamma_{\nu} \psi_\rho 
\nonumber\\
&{}\,
+\frac1{\eta}\,\varepsilon^{\mu\nu\rho}
\, \bar\Psi{}_\mu D[\varpi]_\nu \Psi{}_\rho
  -\frac{\tau}2\,
  \varepsilon^{\mu\nu\rho}
 \,\bar\Psi{}_\mu \gamma_{\nu} \Psi_\rho 
  +\tau\,\varepsilon^{\mu\nu\rho}
 \,\left(\bar\Psi{}_\mu - \bar\psi{}_\mu \right) \gamma_{\nu} \left(\Psi_\rho - \psi_\rho \right) 
\,,
\label{eq:LMMSG}
\end{align}
where ${\cal L}_{\rm CS}$ is the SO$(2,1)$ Chern-Simons Lagrangian for the connection $\varpi_{\mu}{}^{a}$
\begin{equation}
{\cal L}_{\rm CS}[\varpi] = 
\varepsilon^{\mu\nu\rho}
\left(
\varpi_\mu{}^{a} \partial_\nu \varpi_{\rho a}
+\frac{1}{3}\varepsilon_{a b c}\, \varpi_\mu{}^{a} \varpi_\nu{}^{b} \varpi_\rho{}^{c}
\right) \,.
\label{eq:LCS}
\end{equation}
The Lagrangian \eqref{eq:LMMSG} should be viewed as a second order Lagrangian in $e_{\mu}{}^{a}$, with the spin connection $\omega_{\mu}{}^{a}$ dependent on $e_{\mu}{}^{a}$ and $\psi_\mu$ in the following way:
\begin{align}
\omega_\mu{}^a =&\,
\mathring{\omega}_\mu{}^a 
-\tfrac{1}4\,\epsilon^{\rho\sigma\tau}\, e_\rho{}^{a}  \, 
\bar\psi_\sigma \gamma_{\mu} \psi_\tau
+\tfrac{1}8\, \epsilon^{\rho\sigma\tau} \,e_\mu{}^{a}  \,
\bar\psi_\sigma \gamma_{\rho} \psi_\tau
\;,
\label{eq:omega}
\end{align}
where $\mathring{\omega}_\mu{}^a$ is the torsionless Levi-Civita spin connection:
\begin{align} \label{eq:LCconn}
    \mathring{\omega}_{\mu}{}^a &= - \varepsilon^{a b c} e_{\mu d} e_{b}{}^{\nu} e_{c}{}^{\rho} \partial_{[\nu} e_{\rho]}{}^d + \frac12 \varepsilon^{b c d} e_{\mu}{}^{a} e_{c}{}^{\nu} e_{d}{}^{\rho} \partial_{[\nu} e_{\rho] b} \,.
\end{align}
The curvature of $\omega_{\mu}{}^a$ and torsion of $\varpi_{\mu}{}^{a}$ that appear in the first line of \eqref{eq:LMMSG} are defined by
\begin{align}
    R[\omega]_{\mu\nu}{}^a \coloneqq&\, 2\partial_{[\mu} \omega_{\nu]}{}^a + {\varepsilon^a}_{b c} \omega_\mu{}^b \omega_\nu{}^c \,, \nonumber \\
    T[\varpi]_{\mu\nu}{}^{a} \coloneqq&\,2 D[\varpi]_{[\mu} e_{\nu]}{}^a =\,
2\,\partial _{[\mu} e_{\nu]}{}^a + 2\,{\varepsilon^a}_{bc} \,\varpi_{[\mu}{}^{b} e_{\nu]}{}^{c} \,,
\end{align}
while spinor covariant derivatives are defined as
\begin{equation}
D[\omega]_\mu \epsilon =  \partial_\mu \epsilon+\tfrac12\omega_\mu{}^a \gamma_a\,\epsilon \;,\quad
D[\varpi]_\mu \epsilon  =  \partial_\mu \epsilon+\tfrac12\varpi_\mu{}^a \gamma_a\,\epsilon
\;.
\end{equation}
The Lagrangian \eqref{eq:LMMSG} furthermore depends on four parameters $\lambda$, $\tau$, $\kappa$ and $\eta$ and it is assumed that $\kappa \tau \neq 0$. Actually, only three of these parameters are independent, since there exists the following relation among them
\begin{equation}
    \lambda = \frac{1}{12} \left(\eta \tau + \frac{1}{\eta \kappa}\right)^2 - \frac{\tau}{3} \left(\eta \tau - \frac{1}{\eta \kappa}\right) \,.
    \label{lambda}
\end{equation}
Up to quartic fermion terms, the Lagrangian \eqref{eq:LMMSG} is invariant under the following supersymmetry transformation rules:
\begin{align}
\delta_\epsilon e_\mu{}^{a} =\;&
\frac{1}{2}\,\bar\psi_\mu \gamma^{a} \epsilon
\;,\nonumber\\
\delta_\epsilon \psi_\mu =\;& D[\omega]_\mu \epsilon 
-\frac14 \left(\eta\tau+\tfrac1{\eta\,\kappa}\right) \gamma_\mu \epsilon
\;,\nonumber\\
\delta_\epsilon \varpi_\mu{}^{a} =\;&
-\frac1{2\eta\kappa}\,\bar\Psi_\mu  \gamma^{a} \epsilon
-\frac{1}{2}\,D[\varpi]_\mu \left(\bar\Psi_\nu \epsilon \,e^{\nu a} - \bar\psi_\nu \epsilon \,e^{\nu a}\right)
,\nonumber\\
\delta_\epsilon \Psi_\mu =\;&
D[\varpi]_\mu \epsilon -\frac12\,\eta\tau\, \gamma_\mu \epsilon
+\frac{1}{4}\,(\bar\Psi_\lambda \epsilon - \bar\psi_\lambda \epsilon) \,\gamma^{\lambda}\,\Psi_\mu
\,.
\label{eq:susy_full}
\end{align}

MMSG describes the propagation of a single massive spin 2 mode, along with a fermionic superpartner of spin 3/2. Remarkably, when considered around an AdS vacuum, there exists a region in its parameter space where MMSG is unitary, in the sense that the massive modes that are propagated are not tachyonic, nor ghost-like and that the central charges of its asymptotic Virasoro $\times$ Virasoro symmetry algebra are both positive. For supersymmetric AdS vacua, this unitarity region can, after changing the overall sign of the Lagrangian \eqref{eq:LMMSG}, be succinctly described by the conditions:
\begin{equation} \label{eq:unitarity}
    \eta < 0 \,, \qquad \qquad \qquad \kappa \tau < 0 \,.
\end{equation}
We refer to \cite{Deger:2022gim,Deger:2023eah} for a detailed analysis of the unitarity region around arbitrary, not necessarily supersymmetric AdS vacua.

In the following we will be interested in supersymmetric solutions of MMSG, with vanishing fermions. Let us therefore comment on the bosonic equation of motion of the model. The bosonic part of the Lagrangian \eqref{eq:LMMSG} is at most second order in derivatives for two fields $e_{\mu}{}^{a}$ and $\varpi_{\mu}{}^{a}$. In the equations of motion, the latter field can however be eliminated to yield a third-order bosonic equation of motion for the metric field, 
known as a `third-way consistent' equation, which cannot be derived from an action solely expressed in terms of the metric. 
Specifically, from variation w.r.t.\ $e_{\mu}{}^{a}$, one finds the following solution for $\varpi_{\mu}{}^{a}$ 
\begin{equation} \label{eq:solvarpi}
    \varpi_{\mu}{}^{a} = \mathring{\omega}_\mu{}^a - \frac{1}{\tau} \left(S_{\mu \nu} e^{\nu a} + \frac{3 \lambda}{2} e_{\mu}{}^{a} \right) \,,
\end{equation}
in terms of the Schouten tensor $S_{\mu\nu}$. Using this in the equation of motion of $\varpi_{\mu}{}^{a}$ then leads to the following equation for the metric
    \begin{equation}
\left(1+\frac{3\lambda}{2\tau^2}\right)G_{\mu\nu}-\left(\frac{\tau}{\kappa}+\frac{9\lambda^2}{4\tau^2}\right)g_{\mu\nu} - \frac{1}{\tau}C_{\mu\nu} =\frac{1}{\tau^2} J_{\mu\nu}\,,
\label{eq:MMGh}
\end{equation}
where $G_{\mu\nu}$ is the Einstein tensor, $C_{\mu\nu}$ the Cotton tensor and we have denoted
\begin{equation}\label{J}
J_{\mu\nu}\coloneqq-\frac{1}{2}\epsilon_{\mu\kappa\lambda}  \epsilon_{\nu\sigma\tau} S^{\kappa\sigma} S^{\lambda\tau}\,.
\end{equation}
By performing the parameter redefinitions
\begin{equation} \label{eq:param}
\bar{\sigma} = -\frac{\tau}{\mu} \left(1+\frac{3\lambda}{2\tau^2}\right)\,,
\quad\quad
\bar{\Lambda}_0 = \frac{\tau}{\mu} \left(\frac{\tau}{\kappa}+\frac{9\lambda^2}{4\tau^2}\right) \,,
\quad\quad
\gamma = \frac{\mu}{\tau} \,,
\end{equation}
the equation of motion \eqref{eq:MMGh} yields the original form of the MMG equation \cite{Bergshoeff:2014pca}
\begin{equation}
\bar\sigma\,G_{\mu\nu}
+ \bar\Lambda_0\, g_{\mu\nu} + \frac{1}{\mu} C_{\mu\nu}
= 
-\frac{\gamma}{\mu^2} J_{\mu\nu}\,.
\label{eq:MMG0}
\end{equation}

In order to investigate supersymmetry of our solutions, we will need the Killing spinor equations that result from setting the supersymmetry variations of the gravitini (along with the gravitini themselves) equal to zero, i.e.
\begin{subequations}
    \begin{align}
    & D[\mathring{\omega}]_\mu \epsilon 
+ \frac{m}{2} \gamma_\mu \epsilon = 0 \,, \label{eq:KS1} \\  & D[\varpi]_\mu \epsilon -\frac12\,\eta\tau\, \gamma_\mu \epsilon = 0 \,, \label{eq:KS2}
\end{align}
\end{subequations}
where we have defined
\begin{equation}
    m = - \frac12 \left(\eta \tau + \frac{1}{\eta \kappa} \right) \,.
    \label{m}
\end{equation}
In what follows, we will work in the formulation of the equations of motion, in which $\varpi_{\mu}{}^{a}$ has been eliminated by means of \eqref{eq:solvarpi} to yield the third-order equation \eqref{eq:MMGh}. The second Killing spinor equation \eqref{eq:KS2} is then implied by the first one \eqref{eq:KS1}, as follows from the identity
\begin{align}
    D[\varpi]_\mu \epsilon -\frac12\,\eta\tau\, \gamma_\mu \epsilon = \left(1 - \frac{m}{2\tau}\right) \left(D[\mathring{\omega}]_\mu \epsilon 
+ \frac{m}{2} \gamma_\mu \epsilon \right) - \frac{1}{2\tau} \gamma_{\nu} \gamma_{\mu} \epsilon^{\nu \rho \sigma} D[\mathring{\omega}]_\rho \left(D[\mathring{\omega}]_\sigma \epsilon 
+ \frac{m}{2} \gamma_\sigma \epsilon \right) \,,
\end{align}
where it is understood that $\varpi_{\mu}{}^{a}$ is given by \eqref{eq:solvarpi}. When analyzing the Killing spinor equations and their consequences, it will thus suffice to only consider the first equation \eqref{eq:KS1}.

For maximally symmetric solutions of the model for which $G_{\mu\nu}= \Lambda g_{\mu\nu}$, where $\Lambda$ is the cosmological constant, the field equation \eqref{eq:MMGh} implies
\begin{align} \label{cc}
    \Lambda^2 + (4\tau^2 +6\lambda)\Lambda + \left(9\lambda^2 + \frac{4\tau^3}{\kappa}\right) = 0 \, ,
\end{align}
whose roots are 
\begin{align} \label{cc2}
   \Lambda_{\rm susy}=-m^2 \, , \quad  \Lambda_{\rm ns}=-m^2 + 2\tau  \left((\eta-2)\tau + \frac{1}{\eta \kappa} \right) \, .
\end{align}
Only the AdS (or Minkowski) background corresponding to $\Lambda_{\rm susy}$ preserves supersymmetry \eqref{eq:KS1}. Chiral points of the theory are those for which one of the two central charges of the dual theory vanishes. For the supersymmetric AdS vacuum 
this happens when \cite{Deger:2023eah}:
\begin{align} \label{chiral}
    (\eta-1)(\eta \kappa \tau +1)= 0 \, .
\end{align}

Let us finally mention that changing the sign of $m$, that is sending $m \rightarrow -m$, flips the supersymmetry parameter $\eta$ as 
\begin{equation}
\eta \rightarrow - \frac{1}{\eta \kappa \tau}\;,
\label{flip}
\end{equation}
in the above solution \cite{Deger:2023eah}.
Also, it is worth pointing out that for $\Lambda_{\rm ns}<0$, there is another pair of roots $\tilde\eta$ of (\ref{lambda}) (related by (\ref{flip}))
such that
\begin{equation}
\Lambda_{\rm ns} = -\tilde{m}^2 = 
 - \frac14 \left(\tilde\eta \tau + \frac{1}{\tilde\eta \kappa} \right)^2
\;.
\label{mtilde}
\end{equation}
In other words, every bosonic MMG model admitting two AdS vacua has four different supersymmetric extensions with fermionic couplings defined by $\eta$, $\tilde\eta$ and their images under (\ref{flip}), respectively, such that each AdS vacuum is supersymmetric in two of the extensions and non-supersymmetric in the other two.

\section{All Supersymmetric Solutions of MMSG}

In order to find general supersymmetric solutions of MMSG, we follow the reasoning of \cite{Gibbons:2008vi} for the case of topologically massive supergravity (TMSG) \cite{Deser:1982sw,Deser:1982sv}. The bosonic part of the TMSG field equation is of the form \eqref{eq:MMG0} with
$\gamma=0$, and its supersymmetry transformation equals \eqref{eq:KS1}. Assuming the existence of a solution\footnote{This solution for $\epsilon$ is taken to be a \emph{commuting} spinor.} of the Killing spinor equation \eqref{eq:KS1}, one can construct the vector 
\begin{equation}
    K^\mu = \bar{\epsilon} \gamma^\mu \epsilon \,.
\end{equation}
The Fierz identity then implies that $K^\mu$ is a null vector:
\begin{equation}
    K^{\mu} K_{\mu} = 0 \,.
\end{equation}
From \eqref{eq:KS1} it also readily follows that 
\begin{equation} \label{eq:nablaK}
    \nabla_{\mu} K_{\nu} = m \epsilon_{\mu \nu \rho} K^{\rho} \qquad \qquad \Longleftrightarrow \qquad \qquad \epsilon^{\mu\nu\rho} \partial_{\nu} K_{\rho} = - 2 m K^{\mu} \,,
\end{equation}
so that in particular $\nabla_{(\mu} K_{\nu)} = 0$ and one thus concludes that $K^\mu$ is a null Killing vector. The integrability conditions for the Killing spinor equation \eqref{eq:KS1} are identical with those found in \cite{Gibbons:2008vi}.
Following \cite{Gibbons:2008vi}, one can then choose an adapted coordinate $v$ (i.e., a coordinate $v$ in which $K = \partial/\partial v$) and use \eqref{eq:nablaK} to bring the metric to the following form
\begin{equation} \label{eq:gAnsatz}
    \rmd s^2 = \rmd \rho^2 + 2 \rme^{-2 m \rho} \rmd u \rmd v + h(u,\rho) \rmd u^2 \,.
\end{equation}
For this metric, the non-zero components of the Einstein tensor are given by
\begin{align}
    G_{uu} &= {m}^{2} h-m h^{\prime} - \frac{1}{2} h^{\prime\prime} \,, \qquad G_{uv} = {m}^{2} \rme^{-2m \rho} \,, \qquad G_{\rho\rho} = {m}^{2} \,,
\end{align}
where prime ${}^{\prime}$ denotes a derivative with respect to $\rho$.
The non-zero component of the Cotton tensor reads:
\begin{align}
    C_{uu} &= - {m}^{2} h^\prime - \frac{3}{2}m h^{\prime\prime} - \frac{1}{2} h^{\prime\prime\prime} \,.
\end{align}
For the tensor $J_{\mu\nu}$ \eqref{J}, one finds the following non-zero components:
\begin{align}
    J_{uu} &= \frac{m^2}{4} \left({m}^{2} h - 2 m h^\prime -  h^{\prime\prime}\right) \,, \qquad J_{uv} = \frac14 m^4 \rme^{-2 m \rho} \,, \qquad J_{\rho\rho} = \frac14 m^4 \,.
\end{align}
The $(\rho,\rho)$ or $(u,v)$ components of the bosonic field equations \eqref{eq:MMG0} then reduce to the following relation between the parameters of the model
\begin{equation}
    \bar{\sigma} m^2 + \bar{\Lambda}_0 + \frac{\gamma}{4 \mu^2} m^4 = 0 \,,
\end{equation}
which is an identity, upon using (\ref{lambda}), (\ref{eq:param}), (\ref{m}).
The $(u,u)$ component of \eqref{eq:MMG0} becomes
\begin{equation} \label{eq:heq}
  h^{\prime\prime\prime} + \left(2 m -\tau-\frac{1}{\eta \kappa}\right) h^{\prime\prime}  - 2 m \left(\tau + \frac{1}{\eta \kappa} \right) \, h^\prime = 0 \,.
\end{equation}
Assuming separation of variables, the most general solution for $h$ is then given by
\begin{equation} \label{eq:gensol}
    h(u,\rho) = \rme^{\left(\tau + \frac{1}{\eta \kappa}\right) \rho} f_1(u) + \rme^{-2 m \rho} f_2(u) + f_3(u) \,.
\end{equation}
It was shown in \cite{Gibbons:2008vi} that the last two terms are locally redundant, in the sense that they can be removed by coordinate transformations. However, they may be important globally. These terms can actually be generated from the solution without them by using the Garfinkle-Vachaspati method \cite{Garfinkle:1990jq, Garfinkle:1992zj}.
It requires the solution to possess a null Killing vector $K^\mu$ and amounts to finding two functions $\psi$ and $\Omega$ that satisfy
\begin{equation}
\partial_{[\mu}K_{\nu]} = K_{[\mu}\partial_{\nu]}\ln\Omega \,, \qquad 
K^\mu\partial_\mu\psi=0 \,, \qquad \Box\psi=0 \,.
\label{B0}
\end{equation}
Then, the following metric is another exact solution with the same matter fields
\begin{align}
\hat{g}_{\mu\nu}=g_{\mu\nu}+\Omega\psi K_\mu K_\nu \, .
\end{align}
For our solution, one finds
\begin{equation}
    \Omega= \rme^{-2m\rho} \, , \qquad \psi= f_2(u) +f_3(u) \rme^{2m\rho} \, .
\end{equation}
It is interesting to note that the form of the general supersymmetric 
solution is exactly the same as that of TMSG \cite{Gibbons:2008vi}
despite the extra $J$-term in the MMSG field equation \eqref{eq:MMG0}. 
Taking the TMSG limit  ($\gamma \rightarrow 0 \, , \,\bar{\sigma} \rightarrow \sigma \, , \, m \rightarrow m_0)$ as defined in \cite{Deger:2023eah}, the exponent of  the first term goes to $(\tau + \frac{1}{\eta \kappa}) \rightarrow (-m_0-\sigma\mu)$. 

Wave solutions of the form \eqref{eq:gensol} were found and discussed in bosonic MMG in \cite{Alishahiha:2014dma}. In the above derivation, we however fixed an orientation (by taking $\varepsilon^{012} = -1$ as in \cite{Gibbons:2008vi}), which explicitly appears in \eqref{eq:nablaK}. In MMG, one can express the solution in the opposite orientation as well by replacing the coefficient of the Cotton tensor in the field equation (namely $\mu$ in TMSG and $\tau$ in MMSG)
by minus itself in the solution, as was pointed out in \cite{Gibbons:2008vi}. In analogy to the observation of \cite{Gibbons:2008vi} in TMG, we find here that all MMG solutions of type (\ref{eq:gAnsatz}) with a null Killing vector are supersymmetric in one of the supersymmetric extensions of the model, discussed above in (\ref{mtilde}).

When all $f_i$'s in \eqref{eq:gensol} are constant, this spacetime corresponds to \emph{null z-warped AdS} with 
\begin{equation}\label{z}
z=-\frac1{2m}\left(\tau + \frac{1}{\eta \kappa}\right) = \frac{1+\eta \kappa \tau}{1+\eta^2 \kappa \tau}  \, .
\end{equation}
When $z=2$, i.e.
\begin{equation}
  \tau + \frac{1}{\eta \kappa} = - 4 m \qquad \Longleftrightarrow \qquad \tau (2 \eta - 1) + \frac{1}{\eta \kappa} = 0 \,,
\end{equation}
the solution becomes \emph{null warped AdS}, also known as \emph{Schr\"odinger space-time} \cite{Son:2008ye,Balasubramanian:2008dm,Blau:2009gd}. 
Warped AdS solutions of MMG were also discussed in \cite{Arvanitakis:2014yja}.

There are four special points in parameter space, where \eqref{eq:gensol} is no longer the most general solution of \eqref{eq:heq}. This is because of the fact that some of the exponentials appearing in \eqref{eq:gensol} coincide. These special points are given by:
\begin{enumerate}
    \item $\tau + \frac{1}{\eta \kappa} = - 2 m$ and $m\neq 0$: This implies that $\eta = 1$, which is a chiral point \eqref{chiral}. The most general solution of \eqref{eq:heq} is now given by:
    \begin{equation}
        h(u,\rho) = \rho\, \rme^{-2 m \rho} f_1(u) + \rme^{-2 m \rho} f_2(u) + f_3(u) \,.
    \end{equation}
    These solutions do not lie in the unitarity region \eqref{eq:unitarity}.
    \item $\tau + \frac{1}{\eta \kappa} = 0$ and $m\neq 0$: This corresponds to the chiral point $\eta \kappa \tau = -1$ \eqref{chiral}. The most general solution of \eqref{eq:heq} is then given by:
    \begin{equation}
        h(u,\rho) =  \rho f_1(u) + \rme^{-2 m \rho} f_2(u) + f_3(u) \,.
    \end{equation}
    These solutions do not lie in the unitarity region \eqref{eq:unitarity}.
    \item $m = 0$ and $\eta \kappa \tau \neq -1  $: In this case $\bar{\Lambda}_0=0$ in the main field equation \eqref{eq:MMG0}. At this point, the most general solution of \eqref{eq:heq} reads:
    \begin{equation}
        h(u,\rho) = \rme^{\left(\tau + \frac{1}{\eta\kappa}\right)\rho} f_1(u) + \rho f_2(u) + f_3(u) \,.
    \end{equation}
    The vacuum solution in this case is Minkowski spacetime.
    \item $m=0$ and $\eta \kappa \tau = -1$: In this case $\eta=1$ and  $\bar{\sigma}_0=\bar{\Lambda}_0
    = 0$ in the main field equation \eqref{eq:MMG0}. Here, the most general solution of \eqref{eq:heq} is the following:
    \begin{equation}
        h(u,\rho) = \rho^2 f_1(u) + \rho f_2(u) + f_3(u) \,.
    \end{equation}
    The vacuum solution corresponds to Minkowski spacetime.
\end{enumerate}
In all of these special cases, the last two terms of the solution for $h$ are locally redundant \cite{Gibbons:2008vi,Chow:2009km} and can be generated using the Garfinkle-Vachaspati method \cite{Garfinkle:1990jq, Garfinkle:1992zj} described above.

In order to explicitly find the Killing spinor, we need to evaluate the spin connection \eqref{eq:LCconn}. We choose the following dreibein for the metric \eqref{eq:gAnsatz}:
\begin{align}
    e^{0} &= \rme^{-2 m \rho} \rme^{-\beta} \rmd v \,, \qquad  e^{1} = \rme^{-2 m \rho} \rme^{-\beta} \rmd v + \rme^\beta \rmd u \,, \qquad 
    e^{2} = \rmd r \,,
\end{align}
where $h = \rme^{2 \beta}$. We then find the following spin connection components:
\begin{align}
    \omega^{0} &= m \rme^{- 2 m \rho - \beta} \rmd v - \rme^{\beta} \partial_{r} \beta \, \rmd u \,, \qquad \qquad
    \omega^{1} = m \rme^{- 2 m \rho - \beta} \rmd v - \left(m + \partial_r \beta\right) \rme^{\beta} \rmd u \,, \nonumber \\
    \omega^2 &= \partial_{u} \beta\, \rmd u + \left(m + \partial_{r} \beta\right) \rmd r \,.
\end{align}
Choosing $\gamma_0 = \rmi \sigma_2$, $\gamma_1 = \sigma_1$ and $\gamma_2 = \sigma_3$, where $\sigma_i$ are the Pauli matrices, it is then easily checked that 
\begin{align} \label{Killingspinor}
    \epsilon &= h^{-1/4} \, \rme^{- m \rho}\, \epsilon_0\,,
\end{align}
solves the Killing spinor equation \eqref{eq:KS1} where $\epsilon_0$ is a constant spinor satisfying $(\sigma_1 + i\sigma_2)\epsilon_0=0$. So, in general these solutions are 1/2 supersymmetric except for AdS$_3$ (that is when all $f_i$'s are zero in (\ref{eq:gensol})) which is fully supersymmetric.  

Let us finally note that it would be interesting to review in the context of the present model
the Nester-Witten procedure and the positivity property of the associated physical charges \cite{Gibbons:2008vi,Sezgin:2009dj}.

\section{Null Warped AdS Black Holes}

By performing periodic identifications, one can obtain black holes starting from warped AdS geometries \cite{Anninos:2008fx}. For that purpose, we first do the following re-definitions:
\begin{align}
    r=\rme^{-2m\rho} \, , \quad \ \, t=\frac{v}{m} \, , \quad \, \phi =\frac{u}{m} \, ,
\end{align}
and choose the free functions $f_i$ in $h$ given by \eqref{eq:gensol} such that
\begin{align} \label{hnew}
    h(r)= r^{z} + \beta r + \alpha^2 \, , 
\end{align}
where $\alpha$ and $\beta$ are constants and $z$ is defined in \eqref{z}. Then the metric of the supersymmetric solution \eqref{eq:gAnsatz} becomes
\begin{equation} \label{bhmetric}
    \rmd s^2 = \frac{1}{m^2}\left(
     \frac{\rmd r^2}{4r^2} + 2 r \rmd t \rmd \phi + h(r) \rmd \phi^2 \right) \,.
\end{equation}
Taking the coordinate $\phi$ periodic ($\phi \sim \phi + 2\pi$) and assuming that
\begin{align}\label{cond}
    z \geq 2 \, , \quad \beta \geq \alpha \, , \quad  \beta \geq 0 \, , \quad \alpha \neq 0 \, ,
\end{align}
the metric \eqref{bhmetric} describes a black hole with a horizon at $r=0$ \cite{Anninos:2008fx, Clement:2009ka, Anninos:2010pm} and possesses two commuting Killing vectors $\partial_t$ and $\partial_{\phi}$ that are null and spacelike respectively. The conditions \eqref{cond} are needed in order for it to be physically well-defined \cite{Anninos:2010pm}. These black holes are in the unitary region of the theory \eqref{eq:unitarity} provided that
\begin{align}
    -\frac{1}{\eta^2} < \kappa \tau \leq \frac{1}{\eta\,(1 -2\eta)} \, ,
\end{align}
so that the requirement that $z \geq 2$ is satisfied.

Note that the Killing spinor that we found earlier \eqref{Killingspinor} is now independent of the $\phi$ coordinate due to the choice of the function $h$ in \eqref{hnew} and hence is unaffected by its periodic identification. Therefore, the null warped AdS black hole is a supersymmetric solution of MMSG. This might sound surprising, since this black hole has a non-zero entropy and associated temperature \cite{Anninos:2010pm}. However, this notion of temperature differs from the Hawking temperature, which itself is zero since the surface gravity vanishes:
\begin{align}
    k^\nu \nabla_\nu k^\mu = 0 \,, \quad  \text{with} \ \ \ k = \partial_t \,.
\end{align}
As we go to the near horizon limit $r \rightarrow 0$, $h$ approaches to $\alpha^2$ and we get the so-called self-dual AdS$_3$ as the near horizon geometry \cite{Coussaert:1994tu}. By repeating the computation for the Killing spinor above, one sees that in this limit there are two solutions:
\begin{align}
    \epsilon_1 = \begin{pmatrix}
        \sqrt{r} \\ 0 
    \end{pmatrix}\,, \quad   \epsilon_2 = \begin{pmatrix}
        \frac{2}{|\alpha|} t \sqrt{r} \\ \frac{1}{\sqrt{r}} 
    \end{pmatrix} \,,
\end{align}
such that supersymmetry is enhanced. This is unlike some other types of self-dual geometries considered in the literature \cite{Coussaert:1994tu,Balasubramanian:2003kq,OColgain:2016msw}.

\section*{Acknowledgements}

This work is partially supported by a PHC BOSPHORE, projet No 50765SK and by the Scientific and Technological Research Council of T\"urkiye (T\"ubitak) project 123N953. The work of JR is supported by the Croatian Science Foundation project IP-2022-10-5980 ``Non-relativistic supergravity and applications''. JR and NSD are grateful to the Erwin Schr\"odinger Institute (ESI), Vienna where part of this work was done in the framework of the ``Research in Teams'' Programme.


\providecommand{\href}[2]{#2}\begingroup\raggedright\endgroup

\end{document}